\documentclass[%
 aip,
 amsmath,amssymb,
 reprint,%
]{revtex4-1}

\usepackage{subfigure}

\usepackage{graphicx}
\usepackage{dcolumn}
\usepackage{bm}
\usepackage{color}
\usepackage[normalem]{ulem}

\DeclareRobustCommand{\erase}{\bgroup\markoverwith{\textcolor{red}{\rule[.5ex]{2pt}{1pt}}}\ULon}%

\usepackage[utf8]{inputenc}
\usepackage[T1]{fontenc}
\usepackage{mathptmx}
\usepackage{etoolbox}
\usepackage{subcaption}
\usepackage{footmisc} %

\usepackage{devanagari}

\makeatletter
\def\@email#1#2{%
 \endgroup
 \patchcmd{\titleblock@produce}
  {\frontmatter@RRAPformat}
  {\frontmatter@RRAPformat{\produce@RRAP{*#1\href{mailto:#2}{#2}}}\frontmatter@RRAPformat}
  {}{}
}%
\makeatother

\begin{document}

\preprint{AIP/123-QED}

\title[] {\textcolor{black}{Experimental investigation of lift-up and instability of the viscous
flow induced by a rotating cone-cylinder in an enclosure}}

\author{Rajkamal Sah ({\dn{rAjkml sAh}})}
\affiliation{Department of Aerospace Engineering, Indian Institute of Science, Bangalore.}

\author{Sumit Sunil Tambe\textsuperscript{\textdagger} ({\dn {\7{s}E{m}t \7{s}Enl tA\2b\?}}) }
\altaffiliation{Department of Mechanical Engineering, Indian Institute of Technology, Gandhinagar. \newline
\footnotetext{0}{\textsuperscript{\textdagger} Authors to whom the correspondence shall be addressed: sumit.tambe@iitgn.ac.in; jaggie@iisc.ac.in}}

\author{Gopalan Jagadeesh\textsuperscript{\textdagger} 
({\dn{gopAln jg{d}F{f}}})}
    \affiliation{Department of Aerospace Engineering, Indian Institute of Science, Bangalore.}

\date{\today}

\begin{abstract}

This paper probes into the flow induced by a rotating cone-cylinder model in an enclosure. Two component particle image velocimetry measurements in the symmetry plane reveal that the rotating cone-cylinder causes an outward jet on the cylinder section, which lifts the rotating boundary layers away from the wall. A large-scale counter-rotating vortex pair sets up with its mutual upwash aligned with the lift-up region. Furthermore, the centrifugal instability induces Taylor vortices in the rotating boundary layer, which are convected by the mean flow field and are lifted away from the surface, causing a high standard deviation. The lift-up phenomenon shows two preferred axial locations: below a critical Reynolds number $Re_{b,c}$, the lift-up occurs close to the cone-cylinder junction, and for Reynolds number higher than  $Re_{b,c}$ lift-up is pushed away from the cone-cylinder junction, towards the model base. The value of the critical Reynolds number $Re_{b,c}$ lies within $2 \times 10^3-2.5 \times 10^3$ for the investigated cases.

\end{abstract}

\maketitle

\section{Introduction}

The boundary layer flow on rotating bodies, e.g. spheres, cones, and cylinders, is a widely relevant domain of study in fluid mechanics. 

The study of boundary layer flow over rotating geometries like cones, cylinders, and spheres is fundamental to optimizing performance in diverse engineering systems. Rotating boundary layer flow underpin applications such as turbomachinery (e.g., causing drag in aero-engine blades), wind turbines (e.g., aerodynamic efficiency of rotating blades), drilling and boring tools (e.g., precision hole enlargement in aerospace components) \cite{zhao2023aerodynamic}, and projectile design  \cite{suwono1981laminar}. 
Additionally, rotating cylinders and cones are vital in process engineering (e.g., centrifugal separators for gas-solid systems) and gyroscopic stabilization in navigation system \cite{rott1966boundary}. 
Swirling Fluidized Beds (SFBs) utilize rotating annular geometries to separate particulates from gas streams via centrifugal forces, enhancing efficiency in compact industrial cyclones for emissions control and material recovery. In the food industry, Spinning Cone Columns (SCCs) leverage rotating conical surfaces to enhance mixing and mass transfer in confined spaces, enabling efficient extraction of volatile compounds (e.g., flavors, aromas) from viscous liquids without thermal degradation \cite{tanasheva2020aerodynamic}. Also, rotating systems enhance heat/mass transfer in drying applications (e.g., granular materials) and controlled thermal degradation in chemical reactors\cite{jimenez2025swirling}. Therefore, studying the rotating boundary layer flows has received continuous attention over the past century. 

Several studies have provided fundamental knowledge about the boundary layer flows on isolated objects (e.g. cone/disk, sphere, cylinder) rotating in unbounded quiescent flows. However, practical machines often have rotating objects that are combinations of the fundamental shapes, e.g. cone-cylinder body, rotating in a confined environment---where the boundary layer instability affects the overall flow organization within the confinement. Therefore, this paper focuses on studying the development of viscous flow on cone-cylinder geometry in a confined environment.

The rotation effects are known to cause centrifugal, cross-flow, viscous Coriolis, and absolute instabilities in the boundary layers \citep{alfredsson2024flows}---depending on the geometry and boundary conditions of the flow field. Due to the instability, small perturbations in the flow field amplify and induce coherent vortex structures, which grow until the onset of turbulence. The investigation of flow over rotating bodies began with von Kármán's seminal work in 1921 \cite{von1921uber}, which laid the foundation for understanding laminar and turbulent boundary layers on rotating disks. Von Kármán's similarity solution for the laminar boundary layer on a rotating disk has been a cornerstone for subsequent studies on rotating geometries. The progress on the Von Kármán flows---rotating disk and broad cone boundary layers---has been reviewed in detail by \citet{alfredsson2024flows}. \citet{taylor1923viii} investigated the flow between the two concentric cylinders and revealed that the centrifugal instability induces rings of counter-rotating vortices. Furthermore, visualization by \citet{gregory1955stability} revealed that co-rotating spiral vortices appear on a disk rotating in the quiescent fluid before the turbulence onset.  \citet{kobayashi1980spiral,kobayashi1983boundary} investigated spiral vortices on rotating disks and cones in still fluid, showing that the counter-rotating vortices appear on the slender cone but continuously change to co-rotating vortices with increasing cone angle.  Kohama \cite{kohama1984study} provided detailed visualizations of counter-rotating spiral vortices on a rotating cone, showing their growth.  Subsequently, \citet{kobayashi1980spiral, kobayashi1983boundary, hussain2016centrifugal, garrett2009cross} found that, on slender cones, centrifugal instability induces Taylor vortices but on the broad cones, cross-flow instability induces co-rotating spiral vortices similar to those on a rotating disk (which is a cone with half angle $\psi=90^\circ$).

\begin{figure*}[]
    \centering
    \includegraphics[width=0.75\textwidth]{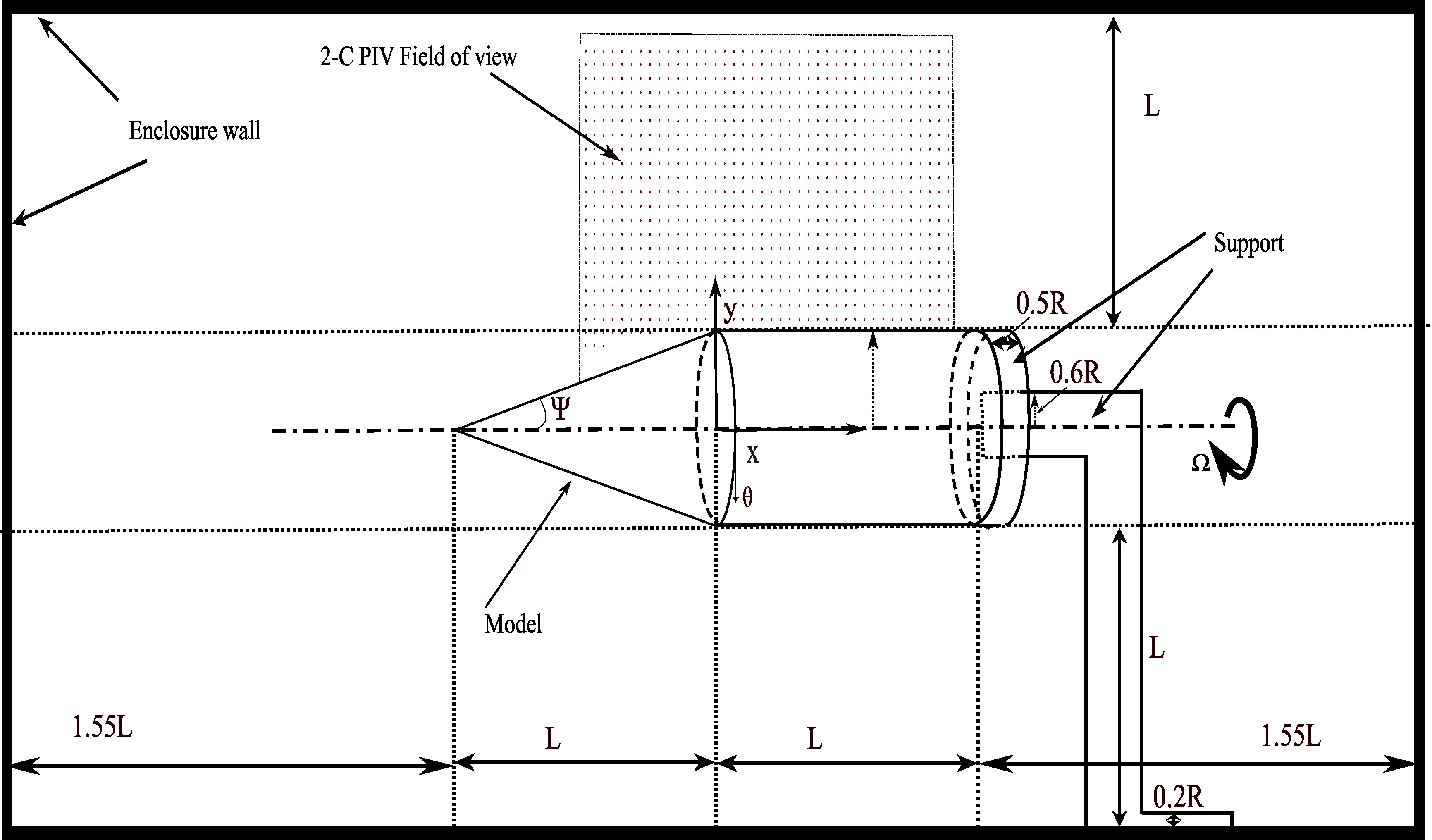}
    \caption{Schematic for experimental setup showing the placement of model in an enclosure and the coordinate system}
    \label{fig:1}
\end{figure*}

Recent studies have shown that the turbulence onset on rotating slender cones ($\psi\leq 30^\circ$) occurs at a well-defined G\"ortler number $G \simeq 10$, in still fluid (\citet{Kato_Segalini_Alfredsson_Lingwood_2021}). Moreover, for a $\psi= 15^\circ$ cone in axial inflow, the boundary layer transition scales with a single parameter, G\"ortler number, regardless of the two different operating parameters oncoming axial flow and rotation rate (\citet{Tambe_Kato_Hussain_2024}).

Miller (1983) \cite{miller1983wind} conducted wind tunnel measurements of the Magnus effect on a spinning projectile, revealing the influence of spin and angle of attack on surface pressure distributions. Sturek \cite{sturek1973boundary} investigated boundary-layer transition on spinning bodies of revolution, emphasizing the role of centrifugal and crossflow instabilities in the transition process. Furthermore, the centrifugal instability modes on rotating cones have been detected and investigated at asymmetric inflow conditions (\citet{Tambe_Schrijer_GangoliRao_Veldhuis_2021,tambe2019experimental}), and at realistic flow conditions for the aero-engines (\citet{Tambe_Realistic_spinner_instability}). 
\cite{}
Chen and Christensen (1967) \cite{chen1967stability} investigated the transition from laminar to turbulent flow in rotating cylinders, identifying a critical Reynolds number for the instability onset. Honji \cite{honji1981streaked} investigated the streaked flow around an oscillating circular cylinder, providing insights into three-dimensional flow instabilities. \citet{MITTAL_KUMAR_2003,Rao_Leontini_Thompson_Hourigan_2013} studied the rotating isolated cylinders in a cross-wise freestream and detected the centrifugal instability modes near the rotating wall. Furthermore, Taylor-Couette \cite{Akira_cylinder_outward_jet,taylor_couette_meeting,taylor1923viii,OutwardJet_freq_lock} and Spherical-Couette \cite{Marcus_Tuckerman_1987,J.P._Sharma_Sameen_Narayanan_2025} flows have been found to exhibit radially outward jets.   

The majority of past studies have investigated the flow on a single type of rotating object (e.g. cone or cylinder) in an ideally-unbounded fluid domain. \textcolor{black}{However, in several practical applications, the rotors are composites of simple geometrical shapes and operate in a confined fluid domain, e.g. stirrers, boring, drilling, process equipment—where studying the rotation-induced flow structures is important. Therefore, in the present work, we probe the flow field around a finite cone-cylinder geometry rotating in a confined fluid domain. The present flow case is different than the well-investigated cases of rotating boundary layers in, ideally unbounded still fluid domain. Apart from influencing the flow around the rotating cone-cylinder, the enclosure contains the seeding particles used for the PIV and shields against any surrounding perturbations that may cause unknown disturbances to the rotating boundary layer system; this ensures that the system under study has well-defined boundary conditions. The experiments done on rotating boundary layers in otherwise still fluid, especially air, need to maintain a high quality of quiet surrounding environment, which---in a practical laboratory setting---translates to a large enclosure lengths as compared to the diameter of the rotating model.} 

Section 2 describes the methodology including the specifications of geometry and experiments. Section 3 presents the results. Section 4 concludes the findings.

\section{Methodology}
The model used in the experiment had a cone-cylinder shape. The half-cone angle $\psi=15^\circ$, which has been the most-investigated case of a slender cone in the literature. The complete model specification and placement of the model inside an enclosed test section are shown in figure \ref{fig:1}. This figure shows the definition of the coordinate system considered in the present investigation. \textcolor{black}{The origin is fixed at the junction of the cone-cylinder. This choice of origin is conveniently separates the representation of two fundamentally different flow cases, i.e. rotating cone and rotating cylinder, which are combined in the present work}. The total length of the cone-cylinder model (2L) is 0.372m which is mounted on a base disk of thickness 0.025m. The base radius\textcolor{black}{, i.e. the radius of the cylinder,} (R) of the model is 0.05m. 

The model was fabricated using CNC machining technology. The surface of the model was black to minimize reflections during laser illumination. The complete model was placed inside a transparent closed box which has dimensions of \(0.95 \, \text{m} \times 0.47 \, \text{m} \times 0.47 \, \text{m}\). The dimensions of the enclosure are sufficiently large so that the boundary layers on the enclosure wall and cone-cylinder wall are separated from each other \textcolor{black}{and the enclosure does not directly influence the development of the rotating-boundary layers.} The rotation rate $\Omega*60/2\pi$ is varied from 72-198 RPM---which is the maximum for the available hardware. \textcolor{black}{Here $\Omega$ is angular velocity.} Table 1 details the investigated cases and \textcolor{black}{the corresponding rotational Reynolds number $Re_b=R^2\Omega/\nu=(R/\delta_\nu)^2$ which compares the azimuthal inertia and viscous effects, here $\nu$ is the kinematic viscosity. By definition, $\sqrt{Re_b}=R/\delta_\nu$, which is nothing but the base radius normalized with the diffusion length scale $\delta_\nu=\sqrt{(\nu/\Omega)}$.} The measured maximum cone-tip eccentricity of the model is around limited to $0.005L$.

\begin{table}
\caption{\label{tab:table2} Measurement matrix.}
\begin{ruledtabular}
\begin{tabular}{cccccccc}
 \textbf{Case} & \textbf{RPM} &$Re_b$ \\
\hline
1 & 198 & 3241 \\
2 & 150 & 2455 \\
3 & 124 & 2030 \\
4 & 98  & 1604 \\
5 & 72 & 1178 \\
\end{tabular}
\end{ruledtabular}

\end{table}

\begin{figure*}[htbp]
    \centering
    \includegraphics[width=1\textwidth,trim={0cm 0cm 0 0cm},clip]{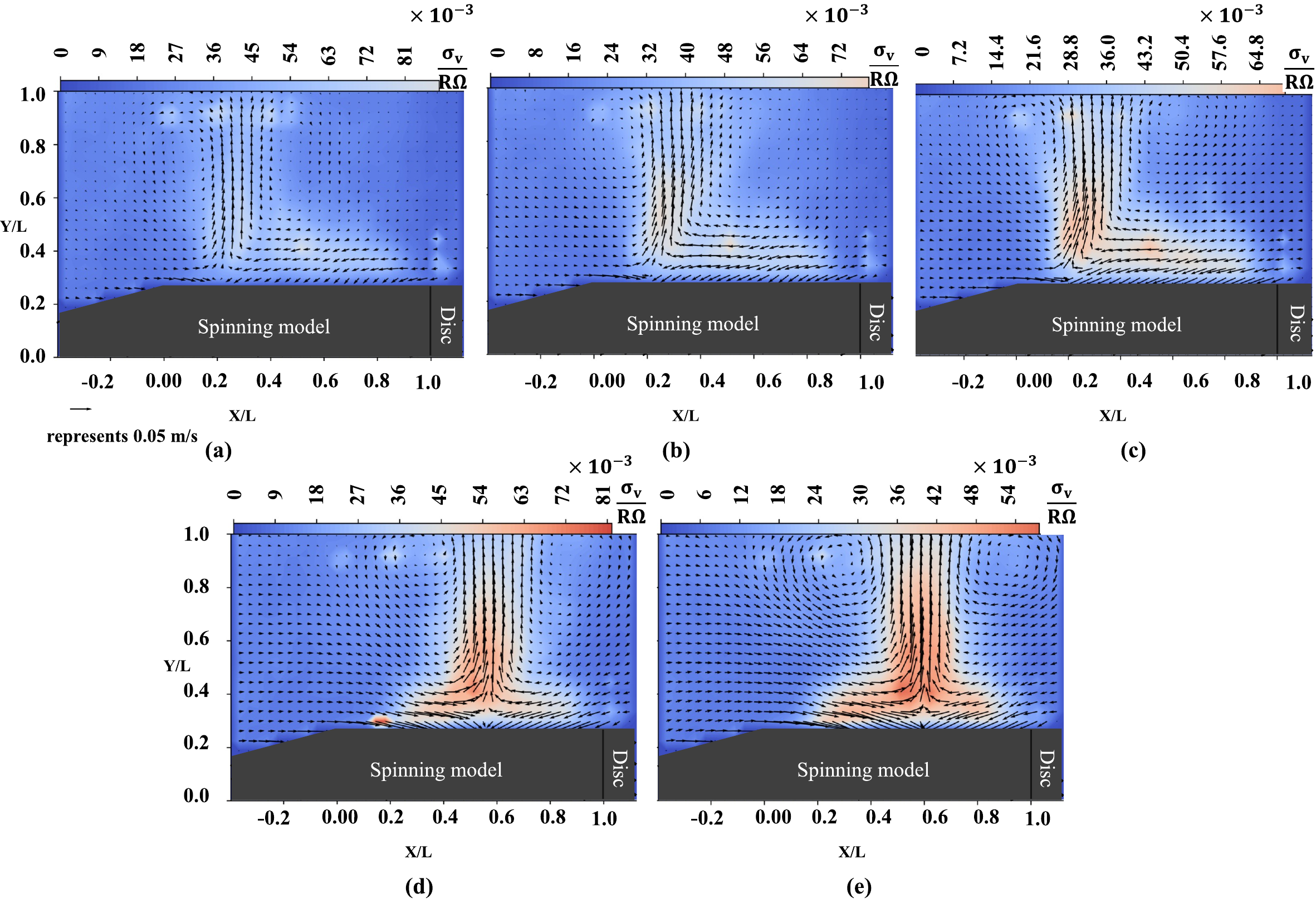}
    \caption{The measured mean velocity vector field and the standard deviation contours in the symmetry plane of the rotating cone-cylinder for different rotation rates: (a) 72 RPM (b) 98 RPM (c) 124 RPM (d) 150 RPM (e) 198 RPM .}
    \label{fig:2}
\end{figure*}

\begin{figure*}[htbp]
    \centering
    \includegraphics[width=1\textwidth,trim={0cm 0cm 0 0cm},clip]{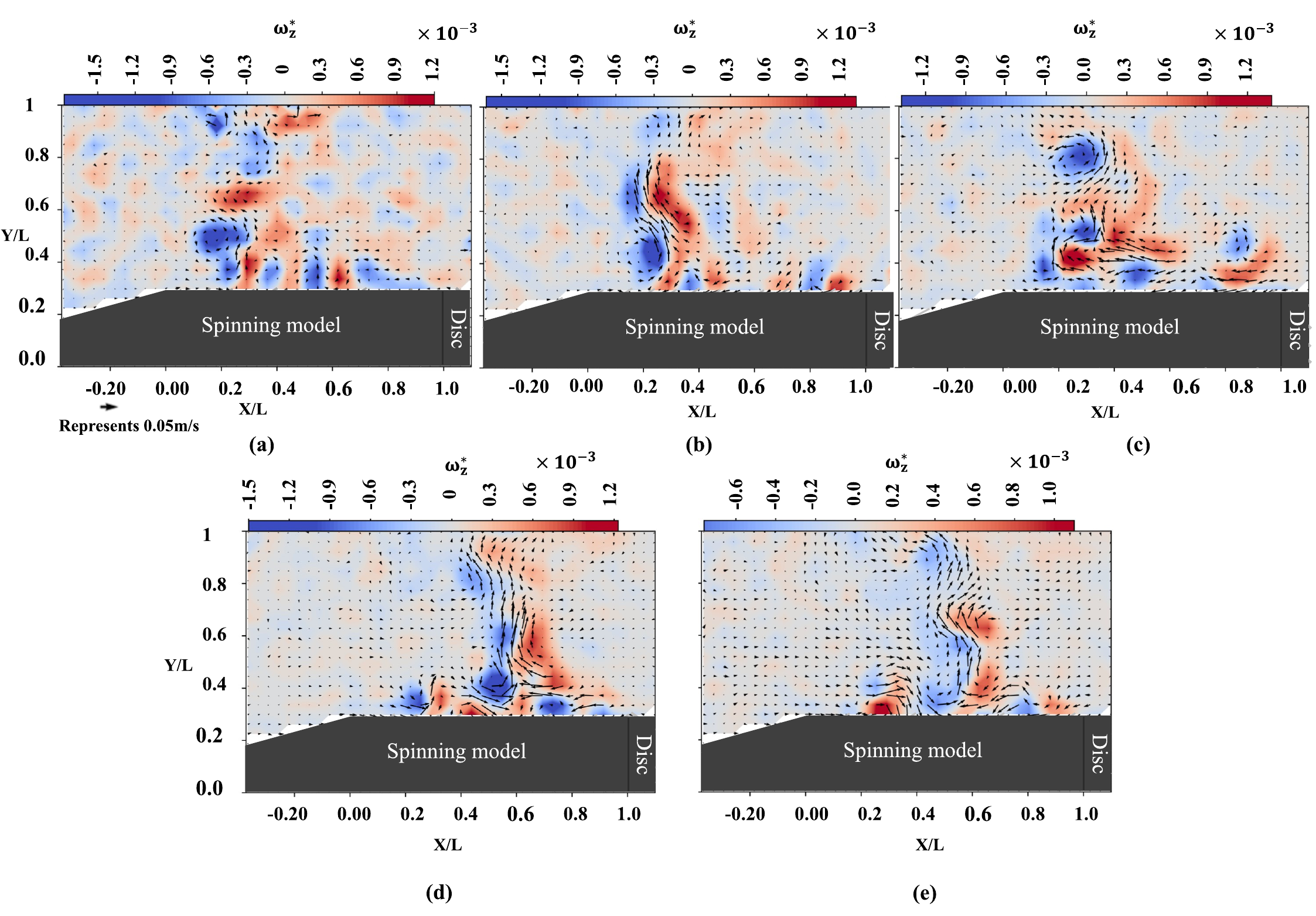}
    \caption{Instantaneous measured vector field and contours of out-of-the-plane vorticity in the symmetry plane of the rotating cone-cylinder for different rotation rates: (a) 72 RPM (multimedia available online) (b) 98 RPM (c) 124 RPM (d) 150 RPM (multimedia available online) (e) 198 RPM.}
    \label{fig:3}
\end{figure*}

Particle Image Velocimetry (PIV) was employed to measure the velocity field of the flow in the symmetry plane of the cone-cylinder model; the field of view is depicted in figure \ref{fig:1}. The flow is seeded with the smoke particles of around $1\mu$m diameter. The experimental setup included a dual-head Nd:YAG laser with a wavelength of \(532 \, \text{nm}\) and a pulse energy of \(135 \, \text{mJ}\) to illuminate the seeding particles. The laser sheet was aligned with the model symmetry plane. MIRO 110 camera with a resolution of \(1280 \times 800\) pixels was positioned perpendicular to the laser sheet, synchronized with the laser pulses to capture particle image pairs at a frequency of \(12.25 \, \text{Hz}\). The image acquisition and post-processing are performed using the commercial code DaVis 8. Cross-correlation is performed on the preprocessed image pairs using a multi-pass approach starting with window size $64 \times 64$ px till $32 \times 32$px. A total of \(600\) image pairs were captured for each test case. The rotation speed is measured by a commercially available tachometer (MEXTECH DT-2234C).

\section{Results}

Figure \ref{fig:2} shows time-averaged vector fields in the symmetry plane of the rotating cone-cylinder model for different rotational rates. The measured symmetry-plane velocity fields show that, close to the model wall, the rotating cone induces flow in a direction towards the base, whereas the flow near the base is directed towards the cone. The two opposing flows meet on the rotating cylinder part and lift the boundary layer up, forming an outward jet in the symmetry plane, similar to the outward jets observed in typical Taylor-Couette \cite{Akira_cylinder_outward_jet} and Spherical-Couette \cite{J.P._Sharma_Sameen_Narayanan_2025} flows.  As the lift-up region extends towards the enclosure wall, it forms a large-scale counter-rotating vortex pair. In figure \ref{fig:2}, the contours represent the normalised standard deviation $\sigma_{v}/R\Omega$ of the axis-normal velocity ($v$) component. The rotating boundary layers on the model and the lift-up region show high values of standard deviation $\sigma_{v}/R\Omega$---suggesting that the rotating-wall-influence extends farther away from the cone-cylinder walls than expected from the laminar boundary layer theory\cite{kobayashi1983boundarystill} for a typical rotating cone. 

\begin{figure*}[htbp]
    \centering
    \includegraphics[width=0.95\textwidth]{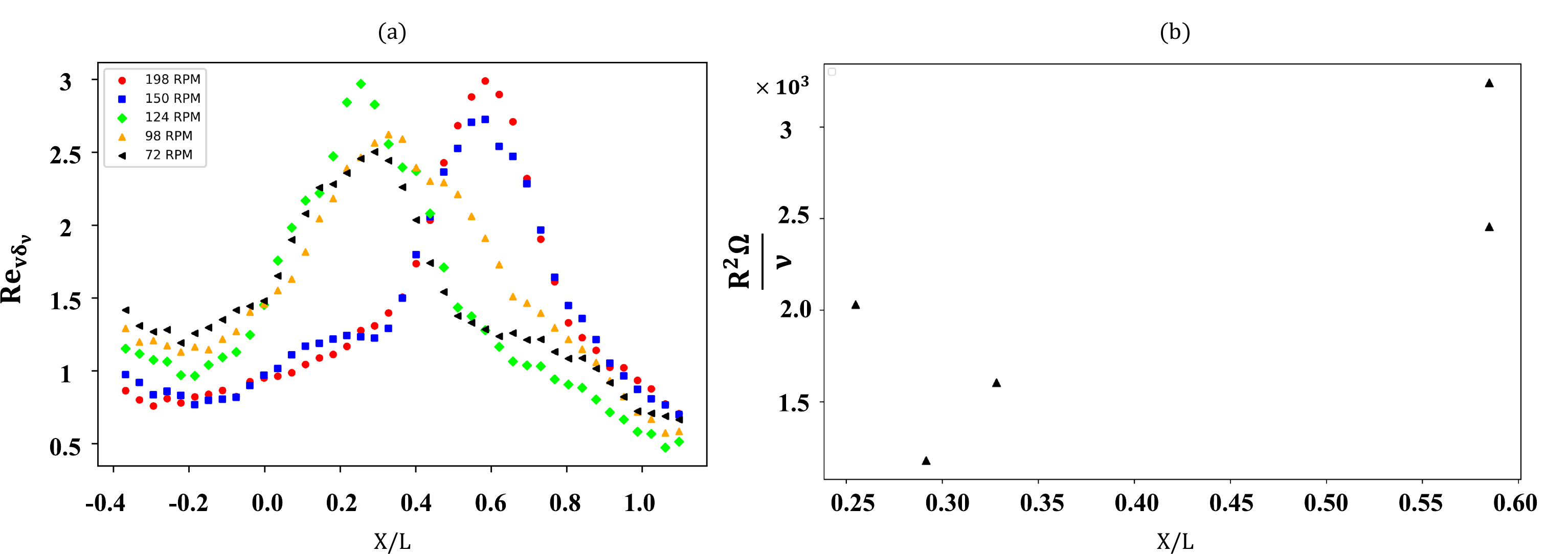}
    \caption{ (a) Reynolds number $Re_{v,\delta_\nu}$ based on the mean axis-normal velocity and the diffusion length scale $\delta_\nu$ traced along the axial direction at $Y/L=0.48$ and (b) the lift-up location in the space of $Re_b$ and $X/L$.}
    \label{fig:9}
\end{figure*}

The lift-up location, i.e. the mutual up-wash of the counter-rotating vortex pair, is observed to depend on the rotational rate (quantified and discussed later with figure \ref{fig:9}). At a low rotation rate, e.g. figure \ref{fig:2}(a), the lift-up region is located just downstream of the cone-cylinder junction. With increasing rotation rate, the cone-induced meridional flow becomes stronger, and the lift-up region shifts farther away from the cone-cylinder junction; see figures \ref{fig:2}(d) and (e). However, all of the investigated cases show that the lift-up phenomena is confined between the end corners of the cylinder.

Figure \ref{fig:3}, shows the instantaneous vector fields corresponding to the respective mean flow fields shown in figure \ref{fig:2}. The contours represent the out-of-the-plane vorticity component $\omega_z^*$. The vorticity contours suggest the existence of the counter-rotating Taylor vortices close to the model wall, caused by the centrifugal instability of the rotating boundary layer. The vortices are convected and lifted-up away from the model wall due to the flow field induced by the rotating cone-cylinder in the enclosure. The convecting Taylor vortices cause the high standard deviation observed in the rotating boundary layer and the lift-up region, shown in figure \ref{fig:2}. At a low rotation rate, e.g. figure \ref{fig:3}(a), the Taylor vortices are observed past the lift-up region, where the train of Taylor vortices is convected towards the cone-cylinder junction and joins the outward jet (see Movie 1 corresponding to \ref{fig:3}(a)). However, at high rotation rates, e.g. figures \ref{fig:3}(d) and (e), the Taylor vortices---appearing on the both sides of the lift-up region---are convected towards each other and mutually interact in the lift-up region (see Movie 2 corresponding to \ref{fig:3}(d)). This interaction is expected to be the cause of the increased standard deviation $\sigma_{v}/R\Omega$ observed in figures \ref{fig:2} (d) and (e).

\begin{figure*}[htbp]
    \centering
    \includegraphics[width=0.95\textwidth]{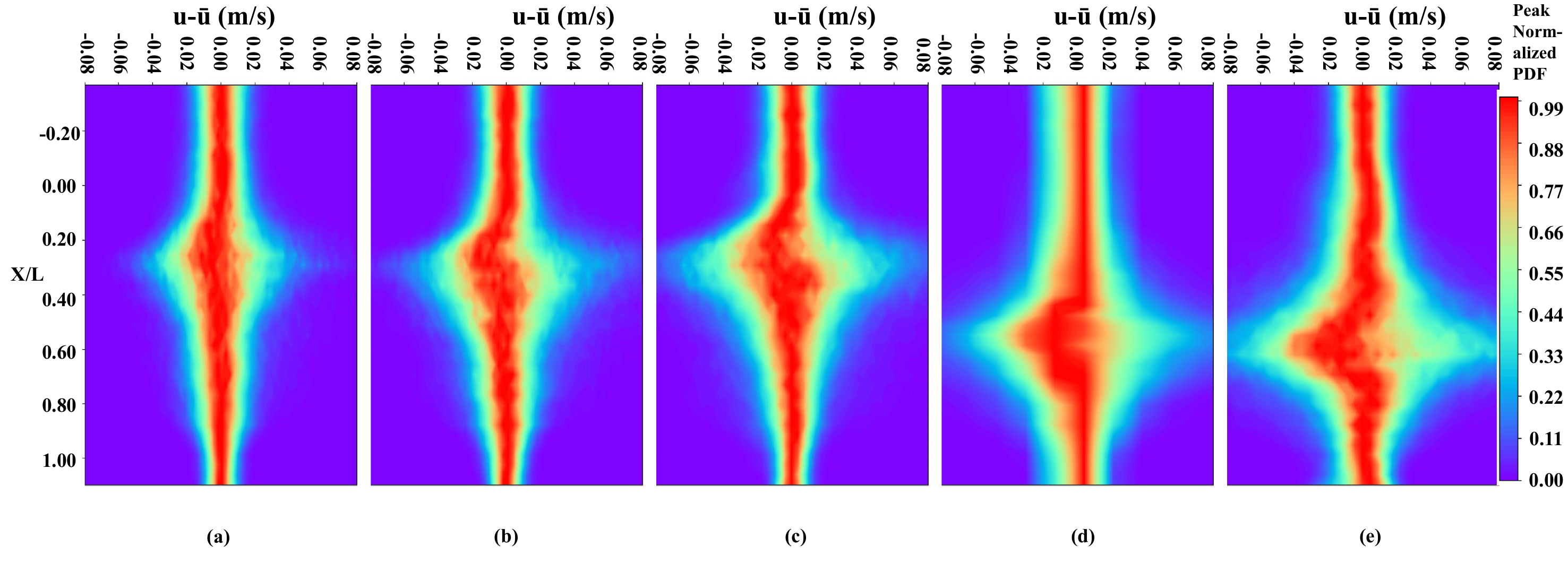}
    \caption{Contours of peak-normalised probability density function for different rotational rates : (a) 72 RPM (b) 98 RPM (c) 124 RPM (d) 150 RPM  (e) 198 RPM.}
    \label{fig:4}
\end{figure*}

The instability-induced Taylor vortices and the rotation-induced lift-up phenomena have a strong influence on the velocity fluctuations in the enclosure. Figure \ref{fig:4} shows distributions of the axial velocity fluctuations at various axial locations, following the procedure similar to that described by \citet{Imayama}. Here, at a given axial location $X/L$, all the measured axis-normal $Y$ locations for 600 images are considered together to obtain a probability density function (p.d.f.) of the axial velocity fluctuations. The p.d.f.s are normalised with their respective peak values and stacked together as per their axial location to obtain the contour plots shown in figure \ref{fig:4}. Upstream of the cone-cylinder junction $X/L<0$, the p.d.f. contours show a symmetric distribution. However, beyond the cone-cylinder junction $X/L>0$, the p.d.f. becomes skewed towards the negative axial velocity fluctuations as it also becomes wider, suggesting the occurrence of extreme events. Beyond the lift-up region, the distribution becomes symmetric again. A similar observation is also reported by \citet{Imayama} where, during the rotating boundary layer transition on a rotating disk, the azimuthal velocity fluctuations are distributed in a skewed fashion and the peak is located towards the negative fluctuations, i.e. slower than the mean velocity. Comparing figures \ref{fig:2}, \ref{fig:3}, and \ref{fig:4} shows that the velocity fluctuations are generally high in the lift-up region, as expected. Furthermore, the widest distribution occurs at two distinct axial locations, i.e. for cases with low rotational rates (below 124 RPM, figure \ref{fig:4}(a)-(c)) the distribution is widest at $X/L\approx 0.3 \pm 0.05$, and for high rotation rates (above 124 RPM, figures \ref{fig:4}(d) and (e)) it occurs at $X/L\approx 0.58$. This suggests that two distinct flow phenomena occur within the range of investigated flow conditions.

\begin{figure*}[htbp]
    \centering
    \includegraphics[width=0.80\textwidth]{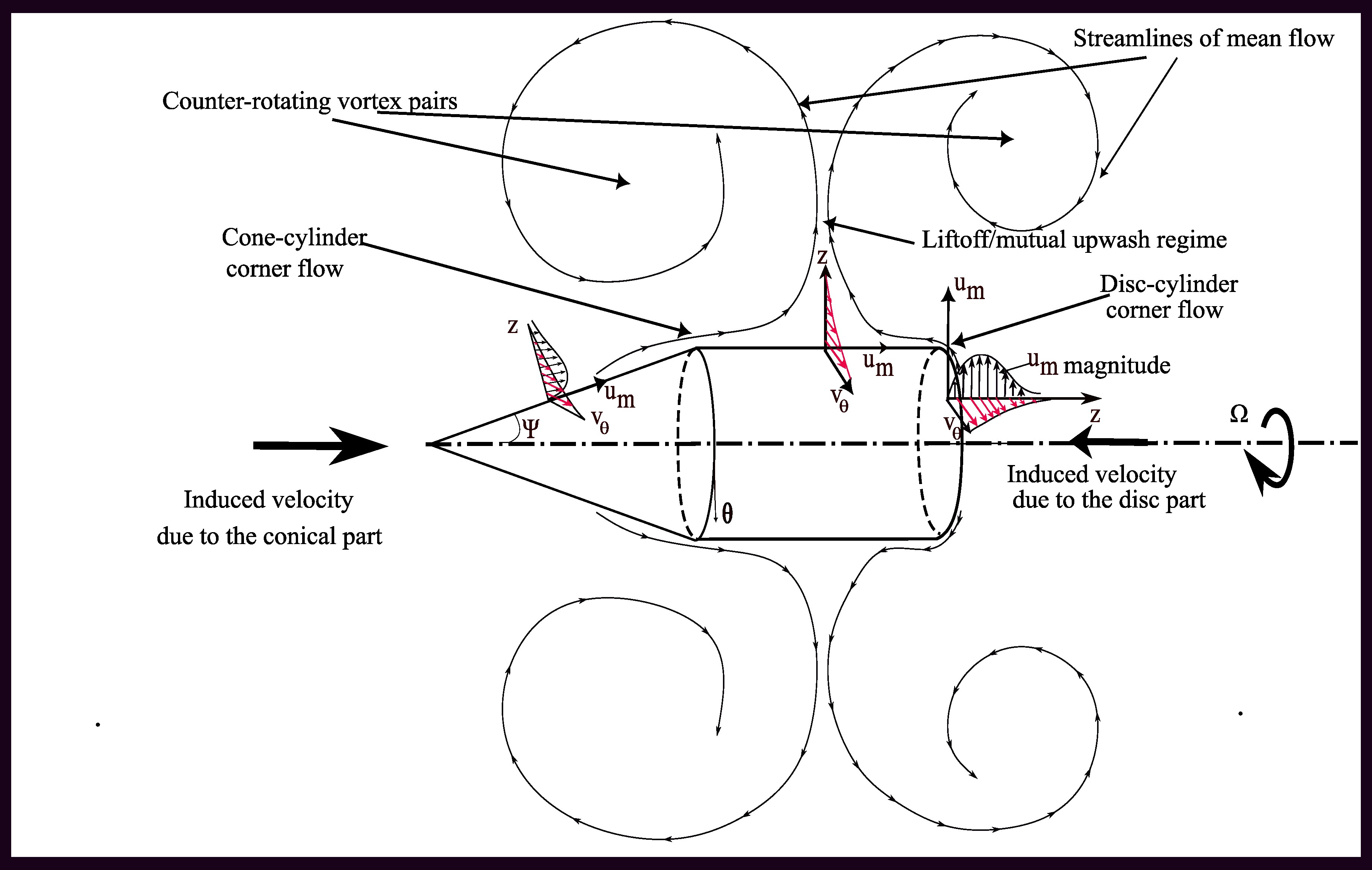}
    \caption{A schematic mean flow organisation around the rotating cone-cylinder in an enclosure.}
    \label{fig:10}
\end{figure*}

The lift-up location is quantified by tracing Reynolds number $Re_{v\delta_{\nu}}=v\delta_{\nu}/\nu$ based on the axis-normal velocity $v$ and the diffusion length scale $\delta_\nu=\sqrt{\nu/\Omega}$, traced at $Y=0.48L$ which is away from the cone-cylinder boundary, see figure \ref{fig:9}a. Profiles of $Re_{v\delta_{\nu}}$ show clear peaks that indicate the mutual upwash of the large-scale counter-rotating vortex pair, and the meridional locations of the peaks are ascribed to the lift-up locations. Figure \ref{fig:9}b shows the lift-up locations in the axial location $X/L$ vs rotational Reynolds number $Re_b=R^2\Omega /\nu$ space. Figure \ref{fig:9}a further highlights the two-distinct lift-up phenomena such that two distinct groups of $Re_{v\delta_{\nu}}$ profiles (below and above the case of 124RPM) are observed to overlap with each other---suggesting two distinct flow phenomena. For $Re_b\lesssim 2\times10^3$, the lift-up location stays confined within $X/L=0.3\pm 0.05$. However, increasing the Reynolds number beyond $Re_b > 2\times10^3$, the lift-up location is pushed at higher $X/L\approx 0.58$, further away from the cone-cylinder corner. It is interesting to note that, as per \citet{kobayashi1983boundarystill}, the boundary layer transition region on a $\psi=15^\circ$ cone rotating in still fluid spans $Re=r^2\Omega/\nu \approx 1\times10^3-4.9\times 10^3$, where $r$ is a local cone radius. In the presence of a meridional inflow, the transition is usually delayed to the higher Reynolds number as compared to the still fluid case \cite{hussain2016centrifugal,Tambe_Schrijer_GangoliRao_Veldhuis_2021}. Furthermore, the boundary layer transition on a rotating cone is known to enhance the mixing of the meridional flow, increasing the near-wall momentum \citet{Tambe_Schrijer_GangoliRao_Veldhuis_2021}. \textcolor{black}{The flow with high momentum is robust against the adverse pressure gradient, which in this case exists past the cone-cylinder junction corner and due to the reverse flow from the cylinder base. Furthermore, the isolated study on a rotating cylinder (\citet{BHORANIYA2024109241}) have shown that increasing meridional velocity stabilizes the boundary layer on an isolated rotating cylinder. Therefore, increasing the Reynolds number beyond a critical $Re_{b,c}$ is expected to cause the boundary layer transition on the rotating cone part, enhancing the near-wall momentum such that the flow over the rotating cone-cylinder junction sustains the adverse pressure gradient for a longer extent and the consequent lift-up location is pushed further towards the base of the model.} For the investigated cases, critical Reynolds number $Re_{b,c}$ lies within $2 \times 10^3-2.5 \times 10^3$. However, this aspect needs further investigation with the close-wall measurements on the rotating cone-cylinder junction.

Conceptually, the lift-up phenomena on the rotating cone-cylinder in an enclosure is a result of two competing pressure gradients set by the two cones at the ends, i.e. a slender cone half angle $\psi=15^\circ$ and a base disk which is a cone with $\psi=90^\circ$---the lift-up mechanism is similar to the outward jet produced in the spherical Couette flows. Figure \ref{fig:10} shows a schematic flow organisation around the rotating cone-cylinder. The pressure gradient on the slender cone draws the flow along its meridian which further accelerates as it crosses the convex-shaped cone-cylinder junction. On the other hand, the radially favorable pressure gradient on the rotating base disk induces axial flow towards the disk, in a direction opposite to that caused by the rotating slender cone. The flow crosses over the convex-shaped rotating cylinder base corner, moves towards the cone-cylinder junction, and interacts with the oncoming cone-induced flow---this potentially causes a lift-up effect observed in the time-averaged flow fields. As observed in figure \ref{fig:3}, the instability-induced Taylor vortices in the rotating boundary layer convect and join the lift-up region, causing high standard deviation.

\section{Conclusion}

When considered separately, both a slender rotating cone and a rotating cylinder are known to induce counter-rotating Taylor vortices caused by the centrifugal instability of their respective boundary layers. As several practical rotating components often have a cone-cylinder shape, i.e. projectiles, stirrers, engine hubs, drilling or boring tools, etc., the present study has explored the flow organization around a combined cone-cylinder body rotating in an enclosure, diagnosed using 2C Particle Image Velocimetry. Following are the conclusions: 

\begin{itemize}  
    
    \item  The viscous flows induced by the rotating cone, cylinder, and base disk interact to cause a lift-up phenomena i.e. an outward jet in the symmetry plane, where the viscous boundary layer is lifted away from the cylinder wall. Therefore, in an enclosed field, the viscous flow around the rotating cone-cylinder body extends far beyond their respective boundary-layer extents known from the isolated cones or cylinder measurements. 
    
    \item The centrifugal instability-induced Taylor vortices appear in the rotating boundary layer and unsteadily interact during the lift-up region, causing high standard deviation.
    \item Two distinct groups of lift-up locations appear: 1) for Reynolds number lower than the critical $Re_{b,c}$, the lift-up is confined close to the cone-cylinder junction $X/L=0.3 \pm 0.05$ and 2) for Reynolds number higher than the critical $Re_{b,c}$, the lift-up is pushed away from the cone-cylinder junction and occurs at $X/L \approx 0.58$. For the investigated cases, critical $Re_{b,c}$ lies within $2 \times 10^3-2.5 \times 10^3$. The boundary layer transition onset on rotating cone (reported by \citet{kobayashi1983boundarystill} to be around $Re=1 \times 10^3 -4.9 \times 10^3$) is expected to play a role in affecting the cone-cylinder corner flow and delaying the lift-up, however, further close-wall investigation is needed to further characterise the flow over the rotating cone-cylinder junction flow.
    \item A large-scale counter-rotating vortex pair sets up around the rotating cone-cylinder with its mutual upwash aligned with the lifted-up viscous flow.

\end{itemize}

The observed lift-up, i.e. an outward jet with unsteady viscous flow, on the rotating cone-cylinder is expected to affect the flow mixing and transport phenomena in an enclosure. In a practical application, e.g. stirring, boring, rotating projectile firing, the unsteady lifted-up viscous flow and the large counter-rotating vortex pair will dominate the flow mixing, transport phenomena, and surface effects (drag and heat transfer). The large-scale counter-rotating vortex pair will cause non-uniform mixing within the enclosure, localizing the scalar transport. Furthermore, due to the lift-up effect, the local skin friction and surface heat transfer will be different than those predicted by considering only cone and cylinder alone. However, formulating the scaling laws of the lift-up location needs a further parametric study considering different values of the, enclosure distances from the model, front and rear cone angles, rotational Reynolds number, and the junction corner curvatures.

\section*{Acknowledgement}

The authors wish to acknowledge the funding support by the Government of India through the Defence Research and Development Organisation (DRDO), Centre of Excellence for Hypersonics (CEH), and Prime Minister Research Fellowship (PMRF) for RS.

\section*{Data availability statement}

Data supporting this manuscript is available upon a reasonable request. 

\section*{References}

\bibliography{reference}

\begin{thebibliography}{33}%
\makeatletter
\providecommand \@ifxundefined [1]{%
 \@ifx{#1\undefined}
}%
\providecommand \@ifnum [1]{%
 \ifnum #1\expandafter \@firstoftwo
 \else \expandafter \@secondoftwo
 \fi
}%
\providecommand \@ifx [1]{%
 \ifx #1\expandafter \@firstoftwo
 \else \expandafter \@secondoftwo
 \fi
}%
\providecommand \natexlab [1]{#1}%
\providecommand \enquote  [1]{``#1''}%
\providecommand \bibnamefont  [1]{#1}%
\providecommand \bibfnamefont [1]{#1}%
\providecommand \citenamefont [1]{#1}%
\providecommand \href@noop [0]{\@secondoftwo}%
\providecommand \href [0]{\begingroup \@sanitize@url \@href}%
\providecommand \@href[1]{\@@startlink{#1}\@@href}%
\providecommand \@@href[1]{\endgroup#1\@@endlink}%
\providecommand \@sanitize@url [0]{\catcode `\\12\catcode `\$12\catcode `\&12\catcode `\#12\catcode `\^12\catcode `\_12\catcode `\%12\relax}%
\providecommand \@@startlink[1]{}%
\providecommand \@@endlink[0]{}%
\providecommand \url  [0]{\begingroup\@sanitize@url \@url }%
\providecommand \@url [1]{\endgroup\@href {#1}{\urlprefix }}%
\providecommand \urlprefix  [0]{URL }%
\providecommand \Eprint [0]{\href }%
\providecommand \doibase [0]{http://dx.doi.org/}%
\providecommand \selectlanguage [0]{\@gobble}%
\providecommand \bibinfo  [0]{\@secondoftwo}%
\providecommand \bibfield  [0]{\@secondoftwo}%
\providecommand \translation [1]{[#1]}%
\providecommand \BibitemOpen [0]{}%
\providecommand \bibitemStop [0]{}%
\providecommand \bibitemNoStop [0]{.\EOS\space}%
\providecommand \EOS [0]{\spacefactor3000\relax}%
\providecommand \BibitemShut  [1]{\csname bibitem#1\endcsname}%
\let\auto@bib@innerbib\@empty
\bibitem [{\citenamefont {Zhao}\ \emph {et~al.}(2023)\citenamefont {Zhao}, \citenamefont {Zhang}, \citenamefont {Bi}, \citenamefont {Zheng}, \citenamefont {Zhong},\ and\ \citenamefont {Zhang}}]{zhao2023aerodynamic}%
  \BibitemOpen
  \bibfield  {author} {\bibinfo {author} {\bibfnamefont {D.}~\bibnamefont {Zhao}}, \bibinfo {author} {\bibfnamefont {Y.}~\bibnamefont {Zhang}}, \bibinfo {author} {\bibfnamefont {M.}~\bibnamefont {Bi}}, \bibinfo {author} {\bibfnamefont {X.}~\bibnamefont {Zheng}}, \bibinfo {author} {\bibfnamefont {X.}~\bibnamefont {Zhong}}, \ and\ \bibinfo {author} {\bibfnamefont {S.}~\bibnamefont {Zhang}},\ }\bibfield  {title} {\enquote {\bibinfo {title} {The aerodynamic characteristics of a rotating cylinder based on large-eddy simulations},}\ }\href@noop {} {\bibfield  {journal} {\bibinfo  {journal} {Journal of Marine Science and Engineering}\ }\textbf {\bibinfo {volume} {11}},\ \bibinfo {pages} {1162} (\bibinfo {year} {2023})}\BibitemShut {NoStop}%
\bibitem [{\citenamefont {Suwono}(1981)}]{suwono1981laminar}%
  \BibitemOpen
  \bibfield  {author} {\bibinfo {author} {\bibfnamefont {A.}~\bibnamefont {Suwono}},\ }\bibfield  {title} {\enquote {\bibinfo {title} {Laminar boundary layer flows near rotating bodies of revolution of arbitrary contour},}\ }\href@noop {} {\bibfield  {journal} {\bibinfo  {journal} {Acta Mechanica}\ }\textbf {\bibinfo {volume} {39}},\ \bibinfo {pages} {51--63} (\bibinfo {year} {1981})}\BibitemShut {NoStop}%
\bibitem [{\citenamefont {Rott}\ and\ \citenamefont {Lewellen}(1966)}]{rott1966boundary}%
  \BibitemOpen
  \bibfield  {author} {\bibinfo {author} {\bibfnamefont {N.}~\bibnamefont {Rott}}\ and\ \bibinfo {author} {\bibfnamefont {W.}~\bibnamefont {Lewellen}},\ }\bibfield  {title} {\enquote {\bibinfo {title} {Boundary layers in rotating flows},}\ }in\ \href@noop {} {\emph {\bibinfo {booktitle} {Applied Mechanics: Proceedings of the Eleventh International Congress of Applied Mechanics Munich (Germany) 1964}}}\ (\bibinfo {organization} {Springer},\ \bibinfo {year} {1966})\ pp.\ \bibinfo {pages} {1030--1036}\BibitemShut {NoStop}%
\bibitem [{\citenamefont {Tanasheva}\ \emph {et~al.}(2020)\citenamefont {Tanasheva}, \citenamefont {Chirkova}, \citenamefont {Dyusembaeva},\ and\ \citenamefont {Sadenova}}]{tanasheva2020aerodynamic}%
  \BibitemOpen
  \bibfield  {author} {\bibinfo {author} {\bibfnamefont {N.}~\bibnamefont {Tanasheva}}, \bibinfo {author} {\bibfnamefont {L.}~\bibnamefont {Chirkova}}, \bibinfo {author} {\bibfnamefont {A.}~\bibnamefont {Dyusembaeva}}, \ and\ \bibinfo {author} {\bibfnamefont {K.}~\bibnamefont {Sadenova}},\ }\bibfield  {title} {\enquote {\bibinfo {title} {Aerodynamic characteristics of a rotating cylinder in the form of a truncated cone},}\ }\href@noop {} {\bibfield  {journal} {\bibinfo  {journal} {Journal of Engineering Physics and Thermophysics}\ }\textbf {\bibinfo {volume} {93}},\ \bibinfo {pages} {551--555} (\bibinfo {year} {2020})}\BibitemShut {NoStop}%
\bibitem [{\citenamefont {Jim{\'e}nez}\ \emph {et~al.}(2025)\citenamefont {Jim{\'e}nez}, \citenamefont {Verdeza}, \citenamefont {Orozco-Jimenez}, \citenamefont {Bula}, \citenamefont {Perreault},\ and\ \citenamefont {Gonzalez-Quiroga}}]{jimenez2025swirling}%
  \BibitemOpen
  \bibfield  {author} {\bibinfo {author} {\bibfnamefont {G.}~\bibnamefont {Jim{\'e}nez}}, \bibinfo {author} {\bibfnamefont {A.}~\bibnamefont {Verdeza}}, \bibinfo {author} {\bibfnamefont {A.~J.}\ \bibnamefont {Orozco-Jimenez}}, \bibinfo {author} {\bibfnamefont {A.}~\bibnamefont {Bula}}, \bibinfo {author} {\bibfnamefont {P.}~\bibnamefont {Perreault}}, \ and\ \bibinfo {author} {\bibfnamefont {A.}~\bibnamefont {Gonzalez-Quiroga}},\ }\bibfield  {title} {\enquote {\bibinfo {title} {Swirling fluidized bed hydrodynamics: Experimental and angular momentum-based assessment},}\ }\href@noop {} {\bibfield  {journal} {\bibinfo  {journal} {Chemical Engineering Journal}\ }\textbf {\bibinfo {volume} {505}},\ \bibinfo {pages} {158867} (\bibinfo {year} {2025})}\BibitemShut {NoStop}%
\bibitem [{\citenamefont {Alfredsson}, \citenamefont {Kato},\ and\ \citenamefont {Lingwood}(2024)}]{alfredsson2024flows}%
  \BibitemOpen
  \bibfield  {author} {\bibinfo {author} {\bibfnamefont {P.~H.}\ \bibnamefont {Alfredsson}}, \bibinfo {author} {\bibfnamefont {K.}~\bibnamefont {Kato}}, \ and\ \bibinfo {author} {\bibfnamefont {R.}~\bibnamefont {Lingwood}},\ }\bibfield  {title} {\enquote {\bibinfo {title} {Flows over rotating disks and cones},}\ }\href@noop {} {\bibfield  {journal} {\bibinfo  {journal} {Annual Review of Fluid Mechanics}\ }\textbf {\bibinfo {volume} {56}},\ \bibinfo {pages} {45--68} (\bibinfo {year} {2024})}\BibitemShut {NoStop}%
\bibitem [{\citenamefont {Von~K{\'a}rm{\'a}n}(1921)}]{von1921uber}%
  \BibitemOpen
  \bibfield  {author} {\bibinfo {author} {\bibfnamefont {T.}~\bibnamefont {Von~K{\'a}rm{\'a}n}},\ }\bibfield  {title} {\enquote {\bibinfo {title} {Uber laminare und turbulente reibung},}\ }\href@noop {} {\bibfield  {journal} {\bibinfo  {journal} {Z. Angew. Math. Mech.}\ }\textbf {\bibinfo {volume} {1}},\ \bibinfo {pages} {233--252} (\bibinfo {year} {1921})}\BibitemShut {NoStop}%
\bibitem [{\citenamefont {Taylor}(1923)}]{taylor1923viii}%
  \BibitemOpen
  \bibfield  {author} {\bibinfo {author} {\bibfnamefont {G.~I.}\ \bibnamefont {Taylor}},\ }\bibfield  {title} {\enquote {\bibinfo {title} {Viii. stability of a viscous liquid contained between two rotating cylinders},}\ }\href@noop {} {\bibfield  {journal} {\bibinfo  {journal} {Philosophical Transactions of the Royal Society of London. Series A, Containing Papers of a Mathematical or Physical Character}\ }\textbf {\bibinfo {volume} {223}},\ \bibinfo {pages} {289--343} (\bibinfo {year} {1923})}\BibitemShut {NoStop}%
\bibitem [{\citenamefont {Gregory}, \citenamefont {Stuart},\ and\ \citenamefont {Walker}(1955)}]{gregory1955stability}%
  \BibitemOpen
  \bibfield  {author} {\bibinfo {author} {\bibfnamefont {N.}~\bibnamefont {Gregory}}, \bibinfo {author} {\bibfnamefont {J.~T.}\ \bibnamefont {Stuart}}, \ and\ \bibinfo {author} {\bibfnamefont {W.}~\bibnamefont {Walker}},\ }\bibfield  {title} {\enquote {\bibinfo {title} {On the stability of three-dimensional boundary layers with application to the flow due to a rotating disk},}\ }\href@noop {} {\bibfield  {journal} {\bibinfo  {journal} {Philosophical Transactions of the Royal Society of London. Series A, Mathematical and Physical Sciences}\ }\textbf {\bibinfo {volume} {248}},\ \bibinfo {pages} {155--199} (\bibinfo {year} {1955})}\BibitemShut {NoStop}%
\bibitem [{\citenamefont {Kobayashi}, \citenamefont {Kohama},\ and\ \citenamefont {Takamadate}(1980)}]{kobayashi1980spiral}%
  \BibitemOpen
  \bibfield  {author} {\bibinfo {author} {\bibfnamefont {R.}~\bibnamefont {Kobayashi}}, \bibinfo {author} {\bibfnamefont {Y.}~\bibnamefont {Kohama}}, \ and\ \bibinfo {author} {\bibfnamefont {C.}~\bibnamefont {Takamadate}},\ }\bibfield  {title} {\enquote {\bibinfo {title} {Spiral vortices in boundary layer transition regime on a rotating disk},}\ }\href@noop {} {\bibfield  {journal} {\bibinfo  {journal} {Acta Mechanica}\ }\textbf {\bibinfo {volume} {35}},\ \bibinfo {pages} {71--82} (\bibinfo {year} {1980})}\BibitemShut {NoStop}%
\bibitem [{\citenamefont {Kobayashi}, \citenamefont {Kohama},\ and\ \citenamefont {Kurosawa}(1983)}]{kobayashi1983boundary}%
  \BibitemOpen
  \bibfield  {author} {\bibinfo {author} {\bibfnamefont {R.}~\bibnamefont {Kobayashi}}, \bibinfo {author} {\bibfnamefont {Y.}~\bibnamefont {Kohama}}, \ and\ \bibinfo {author} {\bibfnamefont {M.}~\bibnamefont {Kurosawa}},\ }\bibfield  {title} {\enquote {\bibinfo {title} {Boundary-layer transition on a rotating cone in axial flow},}\ }\href@noop {} {\bibfield  {journal} {\bibinfo  {journal} {Journal of Fluid Mechanics}\ }\textbf {\bibinfo {volume} {127}},\ \bibinfo {pages} {341--352} (\bibinfo {year} {1983})}\BibitemShut {NoStop}%
\bibitem [{\citenamefont {Kohama}(1984)}]{kohama1984study}%
  \BibitemOpen
  \bibfield  {author} {\bibinfo {author} {\bibfnamefont {Y.}~\bibnamefont {Kohama}},\ }\bibfield  {title} {\enquote {\bibinfo {title} {Study on boundary layer transition of a rotating disk},}\ }\href@noop {} {\bibfield  {journal} {\bibinfo  {journal} {Acta Mechanica}\ }\textbf {\bibinfo {volume} {50}},\ \bibinfo {pages} {193--199} (\bibinfo {year} {1984})}\BibitemShut {NoStop}%
\bibitem [{\citenamefont {Hussain}\ \emph {et~al.}(2016)\citenamefont {Hussain}, \citenamefont {Garrett}, \citenamefont {Stephen},\ and\ \citenamefont {Griffiths}}]{hussain2016centrifugal}%
  \BibitemOpen
  \bibfield  {author} {\bibinfo {author} {\bibfnamefont {Z.}~\bibnamefont {Hussain}}, \bibinfo {author} {\bibfnamefont {S.~J.}\ \bibnamefont {Garrett}}, \bibinfo {author} {\bibfnamefont {S.}~\bibnamefont {Stephen}}, \ and\ \bibinfo {author} {\bibfnamefont {P.~T.}\ \bibnamefont {Griffiths}},\ }\bibfield  {title} {\enquote {\bibinfo {title} {The centrifugal instability of the boundary-layer flow over a slender rotating cone in an enforced axial free stream},}\ }\href@noop {} {\bibfield  {journal} {\bibinfo  {journal} {Journal of Fluid Mechanics}\ }\textbf {\bibinfo {volume} {788}},\ \bibinfo {pages} {70--94} (\bibinfo {year} {2016})}\BibitemShut {NoStop}%
\bibitem [{\citenamefont {Garrett}, \citenamefont {Hussain},\ and\ \citenamefont {Stephen}(2009)}]{garrett2009cross}%
  \BibitemOpen
  \bibfield  {author} {\bibinfo {author} {\bibfnamefont {S.~J.}\ \bibnamefont {Garrett}}, \bibinfo {author} {\bibfnamefont {Z.}~\bibnamefont {Hussain}}, \ and\ \bibinfo {author} {\bibfnamefont {S.}~\bibnamefont {Stephen}},\ }\bibfield  {title} {\enquote {\bibinfo {title} {The cross-flow instability of the boundary layer on a rotating cone},}\ }\href@noop {} {\bibfield  {journal} {\bibinfo  {journal} {Journal of Fluid Mechanics}\ }\textbf {\bibinfo {volume} {622}},\ \bibinfo {pages} {209--232} (\bibinfo {year} {2009})}\BibitemShut {NoStop}%
\bibitem [{\citenamefont {Kato}\ \emph {et~al.}(2021)\citenamefont {Kato}, \citenamefont {Segalini}, \citenamefont {Alfredsson},\ and\ \citenamefont {Lingwood}}]{Kato_Segalini_Alfredsson_Lingwood_2021}%
  \BibitemOpen
  \bibfield  {author} {\bibinfo {author} {\bibfnamefont {K.}~\bibnamefont {Kato}}, \bibinfo {author} {\bibfnamefont {A.}~\bibnamefont {Segalini}}, \bibinfo {author} {\bibfnamefont {P.}~\bibnamefont {Alfredsson}}, \ and\ \bibinfo {author} {\bibfnamefont {R.}~\bibnamefont {Lingwood}},\ }\bibfield  {title} {\enquote {\bibinfo {title} {Instability and transition in the boundary layer driven by a rotating slender cone},}\ }\href {\doibase 10.1017/jfm.2021.216} {\bibfield  {journal} {\bibinfo  {journal} {Journal of Fluid Mechanics}\ }\textbf {\bibinfo {volume} {915}},\ \bibinfo {pages} {R4} (\bibinfo {year} {2021})}\BibitemShut {NoStop}%
\bibitem [{\citenamefont {Tambe}, \citenamefont {Kato},\ and\ \citenamefont {Hussain}(2024)}]{Tambe_Kato_Hussain_2024}%
  \BibitemOpen
  \bibfield  {author} {\bibinfo {author} {\bibfnamefont {S.}~\bibnamefont {Tambe}}, \bibinfo {author} {\bibfnamefont {K.}~\bibnamefont {Kato}}, \ and\ \bibinfo {author} {\bibfnamefont {Z.}~\bibnamefont {Hussain}},\ }\bibfield  {title} {\enquote {\bibinfo {title} {Görtler-number-based scaling of boundary-layer transition on rotating cones in axial inflow},}\ }\href {\doibase 10.1017/jfm.2024.379} {\bibfield  {journal} {\bibinfo  {journal} {Journal of Fluid Mechanics}\ }\textbf {\bibinfo {volume} {987}},\ \bibinfo {pages} {R3} (\bibinfo {year} {2024})}\BibitemShut {NoStop}%
\bibitem [{\citenamefont {Miller}(1983)}]{miller1983wind}%
  \BibitemOpen
  \bibfield  {author} {\bibinfo {author} {\bibfnamefont {M.}~\bibnamefont {Miller}},\ }\bibfield  {title} {\enquote {\bibinfo {title} {Wind tunnel measurements of the magnus induced surface pressures on a spinning projectile in the transonic speed regime},}\ }in\ \href@noop {} {\emph {\bibinfo {booktitle} {Applied Aerodynamics Conference}}}\ (\bibinfo {year} {1983})\ p.\ \bibinfo {pages} {1838}\BibitemShut {NoStop}%
\bibitem [{\citenamefont {Sturek}(1973)}]{sturek1973boundary}%
  \BibitemOpen
  \bibfield  {author} {\bibinfo {author} {\bibfnamefont {W.~B.}\ \bibnamefont {Sturek}},\ }\bibfield  {title} {\enquote {\bibinfo {title} {Boundary-layer studies on spinning bodies of revolution},}\ }\href@noop {} {\  (\bibinfo {year} {1973})}\BibitemShut {NoStop}%
\bibitem [{\citenamefont {Tambe}\ \emph {et~al.}(2021)\citenamefont {Tambe}, \citenamefont {Schrijer}, \citenamefont {Gangoli~Rao},\ and\ \citenamefont {Veldhuis}}]{Tambe_Schrijer_GangoliRao_Veldhuis_2021}%
  \BibitemOpen
  \bibfield  {author} {\bibinfo {author} {\bibfnamefont {S.}~\bibnamefont {Tambe}}, \bibinfo {author} {\bibfnamefont {F.}~\bibnamefont {Schrijer}}, \bibinfo {author} {\bibfnamefont {A.}~\bibnamefont {Gangoli~Rao}}, \ and\ \bibinfo {author} {\bibfnamefont {L.}~\bibnamefont {Veldhuis}},\ }\bibfield  {title} {\enquote {\bibinfo {title} {Boundary layer instability over a rotating slender cone under non-axial inflow},}\ }\href {\doibase 10.1017/jfm.2020.990} {\bibfield  {journal} {\bibinfo  {journal} {Journal of Fluid Mechanics}\ }\textbf {\bibinfo {volume} {910}},\ \bibinfo {pages} {A25} (\bibinfo {year} {2021})}\BibitemShut {NoStop}%
\bibitem [{\citenamefont {Tambe}\ \emph {et~al.}(2019)\citenamefont {Tambe}, \citenamefont {Schrijer}, \citenamefont {Rao},\ and\ \citenamefont {Veldhuis}}]{tambe2019experimental}%
  \BibitemOpen
  \bibfield  {author} {\bibinfo {author} {\bibfnamefont {S.}~\bibnamefont {Tambe}}, \bibinfo {author} {\bibfnamefont {F.}~\bibnamefont {Schrijer}}, \bibinfo {author} {\bibfnamefont {A.~G.}\ \bibnamefont {Rao}}, \ and\ \bibinfo {author} {\bibfnamefont {L.}~\bibnamefont {Veldhuis}},\ }\bibfield  {title} {\enquote {\bibinfo {title} {An experimental method to investigate coherent spiral vortices in the boundary layer over rotating bodies of revolution},}\ }\href@noop {} {\bibfield  {journal} {\bibinfo  {journal} {Experiments in Fluids}\ }\textbf {\bibinfo {volume} {60}},\ \bibinfo {pages} {1--12} (\bibinfo {year} {2019})}\BibitemShut {NoStop}%
\bibitem [{\citenamefont {Tambe}\ \emph {et~al.}(2022)\citenamefont {Tambe}, \citenamefont {Schrijer}, \citenamefont {Veldhuis},\ and\ \citenamefont {Gangoli~Rao}}]{Tambe_Realistic_spinner_instability}%
  \BibitemOpen
  \bibfield  {author} {\bibinfo {author} {\bibfnamefont {S.}~\bibnamefont {Tambe}}, \bibinfo {author} {\bibfnamefont {F.}~\bibnamefont {Schrijer}}, \bibinfo {author} {\bibfnamefont {L.}~\bibnamefont {Veldhuis}}, \ and\ \bibinfo {author} {\bibfnamefont {A.}~\bibnamefont {Gangoli~Rao}},\ }\bibfield  {title} {\enquote {\bibinfo {title} {Spiral instability modes on rotating cones in high-reynolds number axial flow},}\ }\href {\doibase 10.1063/5.0083564} {\bibfield  {journal} {\bibinfo  {journal} {Physics of Fluids}\ }\textbf {\bibinfo {volume} {34}},\ \bibinfo {pages} {034109} (\bibinfo {year} {2022})},\ \Eprint {http://arxiv.org/abs/https://pubs.aip.org/aip/pof/article-pdf/doi/10.1063/5.0083564/16634415/034109\_1\_online.pdf} {https://pubs.aip.org/aip/pof/article-pdf/doi/10.1063/5.0083564/16634415/034109\_1\_online.pdf} \BibitemShut {NoStop}%
\bibitem [{\citenamefont {Chen}\ and\ \citenamefont {Christensen}(1967)}]{chen1967stability}%
  \BibitemOpen
  \bibfield  {author} {\bibinfo {author} {\bibfnamefont {C.}~\bibnamefont {Chen}}\ and\ \bibinfo {author} {\bibfnamefont {D.~K.}\ \bibnamefont {Christensen}},\ }\bibfield  {title} {\enquote {\bibinfo {title} {Stability of flow induced by an impulsively started rotating cylinder},}\ }\href@noop {} {\bibfield  {journal} {\bibinfo  {journal} {Physics of Fluids}\ }\textbf {\bibinfo {volume} {10}},\ \bibinfo {pages} {1845--1846} (\bibinfo {year} {1967})}\BibitemShut {NoStop}%
\bibitem [{\citenamefont {Honji}(1981)}]{honji1981streaked}%
  \BibitemOpen
  \bibfield  {author} {\bibinfo {author} {\bibfnamefont {H.}~\bibnamefont {Honji}},\ }\bibfield  {title} {\enquote {\bibinfo {title} {Streaked flow around an oscillating circular cylinder},}\ }\href@noop {} {\bibfield  {journal} {\bibinfo  {journal} {Journal of Fluid Mechanics}\ }\textbf {\bibinfo {volume} {107}},\ \bibinfo {pages} {509--520} (\bibinfo {year} {1981})}\BibitemShut {NoStop}%
\bibitem [{\citenamefont {Mittal}\ and\ \citenamefont {Kumar}(2003)}]{MITTAL_KUMAR_2003}%
  \BibitemOpen
  \bibfield  {author} {\bibinfo {author} {\bibfnamefont {S.}~\bibnamefont {Mittal}}\ and\ \bibinfo {author} {\bibfnamefont {B.}~\bibnamefont {Kumar}},\ }\bibfield  {title} {\enquote {\bibinfo {title} {Flow past a rotating cylinder},}\ }\href {\doibase 10.1017/S0022112002002938} {\bibfield  {journal} {\bibinfo  {journal} {Journal of Fluid Mechanics}\ }\textbf {\bibinfo {volume} {476}},\ \bibinfo {pages} {303–334} (\bibinfo {year} {2003})}\BibitemShut {NoStop}%
\bibitem [{\citenamefont {Rao}\ \emph {et~al.}(2013)\citenamefont {Rao}, \citenamefont {Leontini}, \citenamefont {Thompson},\ and\ \citenamefont {Hourigan}}]{Rao_Leontini_Thompson_Hourigan_2013}%
  \BibitemOpen
  \bibfield  {author} {\bibinfo {author} {\bibfnamefont {A.}~\bibnamefont {Rao}}, \bibinfo {author} {\bibfnamefont {J.~S.}\ \bibnamefont {Leontini}}, \bibinfo {author} {\bibfnamefont {M.~C.}\ \bibnamefont {Thompson}}, \ and\ \bibinfo {author} {\bibfnamefont {K.}~\bibnamefont {Hourigan}},\ }\bibfield  {title} {\enquote {\bibinfo {title} {Three-dimensionality in the wake of a rapidly rotating cylinder in uniform flow},}\ }\href {\doibase 10.1017/jfm.2013.362} {\bibfield  {journal} {\bibinfo  {journal} {Journal of Fluid Mechanics}\ }\textbf {\bibinfo {volume} {730}},\ \bibinfo {pages} {379–391} (\bibinfo {year} {2013})}\BibitemShut {NoStop}%
\bibitem [{\citenamefont {Kageyama}\ \emph {et~al.}(2004)\citenamefont {Kageyama}, \citenamefont {Ji}, \citenamefont {Goodman}, \citenamefont {Chen},\ and\ \citenamefont {Shoshan}}]{Akira_cylinder_outward_jet}%
  \BibitemOpen
  \bibfield  {author} {\bibinfo {author} {\bibfnamefont {A.}~\bibnamefont {Kageyama}}, \bibinfo {author} {\bibfnamefont {H.}~\bibnamefont {Ji}}, \bibinfo {author} {\bibfnamefont {J.}~\bibnamefont {Goodman}}, \bibinfo {author} {\bibfnamefont {F.}~\bibnamefont {Chen}}, \ and\ \bibinfo {author} {\bibfnamefont {E.}~\bibnamefont {Shoshan}},\ }\bibfield  {title} {\enquote {\bibinfo {title} {Numerical and experimental investigation of circulation in short cylinders},}\ }\href {\doibase 10.1143/JPSJ.73.2424} {\bibfield  {journal} {\bibinfo  {journal} {Journal of the Physical Society of Japan}\ }\textbf {\bibinfo {volume} {73}},\ \bibinfo {pages} {2424--2437} (\bibinfo {year} {2004})},\ \Eprint {http://arxiv.org/abs/https://doi.org/10.1143/JPSJ.73.2424} {https://doi.org/10.1143/JPSJ.73.2424} \BibitemShut {NoStop}%
\bibitem [{\citenamefont {Bühler}\ \emph {et~al.}(1986)\citenamefont {Bühler}, \citenamefont {Coney}, \citenamefont {Wimmer},\ and\ \citenamefont {Zierep}}]{taylor_couette_meeting}%
  \BibitemOpen
  \bibfield  {author} {\bibinfo {author} {\bibfnamefont {K.}~\bibnamefont {Bühler}}, \bibinfo {author} {\bibfnamefont {J.~E.~R.}\ \bibnamefont {Coney}}, \bibinfo {author} {\bibfnamefont {M.}~\bibnamefont {Wimmer}}, \ and\ \bibinfo {author} {\bibfnamefont {J.}~\bibnamefont {Zierep}},\ }\bibfield  {title} {\enquote {\bibinfo {title} {Advances in taylor vortex flow: A report on the fourth taylor vortex flow working party meeting},}\ }\href@noop {} {\bibfield  {journal} {\bibinfo  {journal} {Acta Mechanica}\ }\textbf {\bibinfo {volume} {62}} (\bibinfo {year} {1986})}\BibitemShut {NoStop}%
\bibitem [{\citenamefont {{von Stamm}}, \citenamefont {Buzug},\ and\ \citenamefont {Pfister}(1994)}]{OutwardJet_freq_lock}%
  \BibitemOpen
  \bibfield  {author} {\bibinfo {author} {\bibfnamefont {J.}~\bibnamefont {{von Stamm}}}, \bibinfo {author} {\bibfnamefont {T.}~\bibnamefont {Buzug}}, \ and\ \bibinfo {author} {\bibfnamefont {G.}~\bibnamefont {Pfister}},\ }\bibfield  {title} {\enquote {\bibinfo {title} {Frequency locking in axisymmetric taylor-couette flow},}\ }\href {\doibase https://doi.org/10.1016/0375-9601(94)91279-3} {\bibfield  {journal} {\bibinfo  {journal} {Physics Letters A}\ }\textbf {\bibinfo {volume} {194}},\ \bibinfo {pages} {173--178} (\bibinfo {year} {1994})}\BibitemShut {NoStop}%
\bibitem [{\citenamefont {Marcus}\ and\ \citenamefont {Tuckerman}(1987)}]{Marcus_Tuckerman_1987}%
  \BibitemOpen
  \bibfield  {author} {\bibinfo {author} {\bibfnamefont {P.~S.}\ \bibnamefont {Marcus}}\ and\ \bibinfo {author} {\bibfnamefont {L.~S.}\ \bibnamefont {Tuckerman}},\ }\bibfield  {title} {\enquote {\bibinfo {title} {Simulation of flow between concentric rotating spheres. part 1. steady states},}\ }\href {\doibase 10.1017/S0022112087003069} {\bibfield  {journal} {\bibinfo  {journal} {Journal of Fluid Mechanics}\ }\textbf {\bibinfo {volume} {185}},\ \bibinfo {pages} {1–30} (\bibinfo {year} {1987})}\BibitemShut {NoStop}%
\bibitem [{\citenamefont {J.P.}\ \emph {et~al.}(2025)\citenamefont {J.P.}, \citenamefont {Sharma}, \citenamefont {Sameen},\ and\ \citenamefont {Narayanan}}]{J.P._Sharma_Sameen_Narayanan_2025}%
  \BibitemOpen
  \bibfield  {author} {\bibinfo {author} {\bibfnamefont {A.}~\bibnamefont {J.P.}}, \bibinfo {author} {\bibfnamefont {M.}~\bibnamefont {Sharma}}, \bibinfo {author} {\bibfnamefont {A.}~\bibnamefont {Sameen}}, \ and\ \bibinfo {author} {\bibfnamefont {V.}~\bibnamefont {Narayanan}},\ }\bibfield  {title} {\enquote {\bibinfo {title} {Bifurcations in narrow-gap spherical couette flow},}\ }\href {\doibase 10.1017/jfm.2025.163} {\bibfield  {journal} {\bibinfo  {journal} {Journal of Fluid Mechanics}\ }\textbf {\bibinfo {volume} {1007}},\ \bibinfo {pages} {A64} (\bibinfo {year} {2025})}\BibitemShut {NoStop}%
\bibitem [{\citenamefont {Kobayashi}\ and\ \citenamefont {Izumi}(1983)}]{kobayashi1983boundarystill}%
  \BibitemOpen
  \bibfield  {author} {\bibinfo {author} {\bibfnamefont {R.}~\bibnamefont {Kobayashi}}\ and\ \bibinfo {author} {\bibfnamefont {H.}~\bibnamefont {Izumi}},\ }\bibfield  {title} {\enquote {\bibinfo {title} {Boundary-layer transition on a rotating cone in still fluid},}\ }\href@noop {} {\bibfield  {journal} {\bibinfo  {journal} {Journal of Fluid Mechanics}\ }\textbf {\bibinfo {volume} {127}},\ \bibinfo {pages} {353--364} (\bibinfo {year} {1983})}\BibitemShut {NoStop}%
\bibitem [{\citenamefont {Imayama}, \citenamefont {Alfredsson},\ and\ \citenamefont {Lingwood}(2012)}]{Imayama}%
  \BibitemOpen
  \bibfield  {author} {\bibinfo {author} {\bibfnamefont {S.}~\bibnamefont {Imayama}}, \bibinfo {author} {\bibfnamefont {P.~H.}\ \bibnamefont {Alfredsson}}, \ and\ \bibinfo {author} {\bibfnamefont {R.~J.}\ \bibnamefont {Lingwood}},\ }\bibfield  {title} {\enquote {\bibinfo {title} {A new way to describe the transition characteristics of a rotating-disk boundary-layer flow},}\ }\href {\doibase 10.1063/1.3696020} {\bibfield  {journal} {\bibinfo  {journal} {Physics of Fluids}\ }\textbf {\bibinfo {volume} {24}},\ \bibinfo {pages} {031701} (\bibinfo {year} {2012})},\ \Eprint {http://arxiv.org/abs/https://pubs.aip.org/aip/pof/article-pdf/doi/10.1063/1.3696020/14764123/031701\_1\_online.pdf} {https://pubs.aip.org/aip/pof/article-pdf/doi/10.1063/1.3696020/14764123/031701\_1\_online.pdf} \BibitemShut {NoStop}%
\bibitem [{\citenamefont {Bhoraniya}\ and\ \citenamefont {Narayanan}(2024)}]{BHORANIYA2024109241}%
  \BibitemOpen
  \bibfield  {author} {\bibinfo {author} {\bibfnamefont {R.}~\bibnamefont {Bhoraniya}}\ and\ \bibinfo {author} {\bibfnamefont {V.}~\bibnamefont {Narayanan}},\ }\bibfield  {title} {\enquote {\bibinfo {title} {Global stability analysis of axisymmetric boundary layer on a rotating circular cylinder},}\ }\href {\doibase https://doi.org/10.1016/j.ijheatfluidflow.2023.109241} {\bibfield  {journal} {\bibinfo  {journal} {International Journal of Heat and Fluid Flow}\ }\textbf {\bibinfo {volume} {105}},\ \bibinfo {pages} {109241} (\bibinfo {year} {2024})}\BibitemShut {NoStop}%
\end{thebibliography}%

\end{document}